\documentclass[aps,prl,twocolumn, titlepage,showpacs]{revtex4}

\usepackage{graphicx}

\usepackage{dcolumn}

\usepackage{bm}

\begin{document}

\title{Errors in particle tracking velocimetry with high-speed cameras}

\author{Yan Feng}
\email{yan-feng@uiowa.edu}
\author{J. Goree}
\author{Bin Liu}
\affiliation{Department of Physics and Astronomy, The University
of Iowa, Iowa City, Iowa 52242}

\date{\today}

\begin{abstract}

Velocity errors in particle tracking velocimetry (PTV) are
studied. When using high-speed video cameras, the velocity error
may increase at a high camera frame rate. This increase in
velocity error is due to particle-position uncertainty, which is
one of two sources of velocity errors studied here. The other
source of error is particle acceleration, which has the opposite
trend of diminishing at higher frame rates. Both kinds of errors
can propagate into quantities calculated from velocity, such as
the kinetic temperature of particles or correlation functions. As
demonstrated in a dusty plasma experiment, the kinetic temperature
of particles has no unique value when measured using PTV, but
depends on the sampling time interval or frame rate. It is also
shown that an artifact appears in an autocorrelation function
computed from particle positions and velocities, and it becomes
more severe when a small sampling-time interval is used. Schemes
to reduce these errors are demonstrated.

\end{abstract}

\narrowtext

\maketitle

\section{I.~Introduction}

Particle tracking velocimetry (PTV) is a method to measure
particle velocities~\cite{Feng:07} with video camera recording. In
experiments, the particles are small solid objects that can
scatter enough light to be imaged separately. The velocity is
calculated based on measured positions of particles. Among the
algorithms used to calculate velocities~\cite{Luthi:05,
Hadziavdic:06, Oxtoby:10}, the most common is simply to divide the
difference in a particle's position in two consecutive video
frames by the time interval between the frames~\cite{Feng:07,
Nunomura:05, Feng:08}. Although PTV may allow tracking individual
particles for many frames, for this algorithm it is only necessary
to track for two frames.

Particle tracking velocimetry has been widely used for many years
in topics in various fields, such as cell motion in
biology~\cite{Lima:09}, flow in granular materials~\cite{Jain:02},
and kinetic temperature in dusty plasmas~\cite{Melzer:96,
Nosenko:04, Nunomura:05, Ivanov:05, Fortov:07, Feng:08,
Knapek_thesis}. Small particles of solid matter can be added as
tracers in a gas or liquid to study convection~\cite{Dore:09} and
turbulence~\cite{Luthi:05} in fluid mechanics. Many dynamical
quantities can be calculated using velocities measurements with
PTV, for example velocity profiles in a shear flow~\cite{Jain:02,
Nosenko:04}, mean-square velocity fluctuations~\cite{Feng:08},
velocity distribution functions~\cite{Melzer:96, Nosenko:06,
Fortov:07}, and velocity autocorrelation
functions~\cite{Vaulina:08}. Some of these uses for PTV, such as
velocity profiles, can also be accomplished with particle image
velocimetry (PIV)~\cite{Westerweel:97, Williams:06}. Compared to
PIV, PTV provides a measurement of velocity at the location of a
particle, without requiring an averaging over a grid.

Recently, many experimenters using PTV have taken advantage of the
abundance of high-speed cameras now offered for commercial sale.
They may be unaware, however, that velocities determined using the
PTV method can have errors that become more severe as the camera
frame rate is increased. For example, we show that when
determining kinetic temperature by calculating the mean-square
velocity fluctuation for random motion, the result will have an
exaggerated value that worsens at higher frame rates, due to one
of the two kinds of errors studied here.

\section{II.~velocity errors in PTV}

We identify two kinds of velocity errors in PTV according to their
source. One kind arises from acceleration of the particle during
the time interval between measurements of its position. The other
arises from errors in the position measurements themselves. While
the former is made less severe by using a faster frame rate, the
latter is actually made worse. Therefore, a faster frame rate is
not always best for PTV.

\subsection{A.~Velocity error arising from acceleration}
An error arises from acceleration, whether due to a change in a
particle's speed or its direction. This error occurs for all kinds
of acceleration, for example a particle accelerating along a
straight line, moving in a circle at a steady speed, or colliding
with another particle.

It is impossible to eliminate this error when the motion is
unknown during the sampling interval between frames. The only
information available in PTV are the particle positions determined
at the times that a video image was recorded, i.e., the video
frames.

The size of this error also depends on the algorithm for
calculating velocity from data for the particle positions. The
simplest and most common algorithm is to assume that the particle
moves with zero acceleration between two consecutive frames. In
this two-frame tracking method, the particle is assumed to move in
a straight line at a steady speed between the positions in two
frames. Thus, the velocity is calculated simply as the difference
in positions divided by the time interval between frames. Using
two position measurements ${\bf x}_{\rm meas}$, denoted by $j$ and
$j+1$ separated by a time $\Delta t$, the velocity at a time
halfway between them is calculated as
\begin{equation}\label{2-frame-velocity}
{{\bf V}=\frac{{\bf x}_{j+1, {\rm meas}}-{\bf x}_{j, {\rm
meas}}}{\Delta t}}.
\end{equation}
The camera's frame rate is $1/\Delta t$ if no frames are skipped.
We use upper-case symbols like $V$ and $X$ to indicate quantities
that are {\it computed} from the positions that are measured in a
single frame. If the particle actually undergoes acceleration, for
example if it has a curved trajectory, the simple algorithm of
Eq.~(\ref{2-frame-velocity}) will obviously lead to errors in the
velocity ${\bf V}$. Alternatively, one could use an algorithm
using more than two position measurements, which can sometimes
better account for acceleration. A spline fit~\cite{Luthi:05}
could reduce the error arising from acceleration, but it is more
computationally expensive than two-frame tracking.

To illustrate how errors in ${\bf V}$ arise from acceleration and
how these errors diminish with a higher frame rate, we consider
the motion during a Coulomb collision of a pair of identical
electrically-charged particles. The largest acceleration in this
case will occur when the particles are closest and the direction
of motion is changing most rapidly. This highly curved portion of
a trajectory can be approximated as a circular arc. This motivates
us to choose uniform circular motion as a simple instructive
example of velocity errors arising from acceleration, Fig.~1.
Suppose the particle's position is measured accurately at each
frame, as indicated by the dots in Fig.~1(a). The simple two-frame
tracking method, which assumes zero acceleration between
measurements, will describe the motion as a polygon instead of the
ideal circle. At a fast frame rate, the polygon has more sides and
more closely approximates the ideal circle.

\subsection{B.~Velocity error arising from position uncertainty}

Another error in velocity arises from uncertainties in the
particle positions from which the velocity is computed. Particle
positions are uncertain for at least two reasons: random noise in
the camera sensor, and the finite size of pixels in the sensor.
The latter leads to the phenomenon of pixel locking, where
particles are wrongly identified as being located at favored
positions such as the corner or middle of pixels. While it is
possible to design an experiment to reduce these particle-position
uncertainties~\cite{Feng:07, Knapek_thesis}, they can never be
eliminated.

Suppose that all measured particle positions contain an
uncertainty $\delta {\bf x}$. We can use propagation of errors to
estimate the uncertainty of the calculated velocity ${\bf V}$,
arising from the uncertainties in the particle positions, for any
given algorithm. Choosing the algorithm used in the simple
two-frame tracking method, Eq.~(\ref{2-frame-velocity}), the
uncertainty in the calculated velocity ${\bf V}$ is
\begin{equation}\label{2-frame-error}
{(\delta {\bf V})^2=\frac{(\delta {\bf x}_{j+1})^2+(\delta {\bf
x}_{j})^2}{(\Delta t)^2}=2(\delta {\bf x}/\Delta t)^2.}
\end{equation}

Importantly, this source of error becomes larger, not smaller, as
the time interval between measurements $\Delta t$ is decreased.
This is seen in Eq.~(\ref{2-frame-error}), where the denominator
diminishes with an increasing frame rate but the numerator does
not vary with frame rate at all. Thus, this error in velocity is
$\propto 1/\Delta t$. Now that high frame-rate scientific cameras
have become more easily available, this error in PTV probably
occurs more commonly.

To illustrate the combined effect of both types of errors, those
due to acceleration and due to uncertainties in particle position,
consider the sketch in Fig.~1(b) for a particle undergoing uniform
circular motion. The uncertainty in particle positions is
indicated in Fig.~1(b) by shading around the true positions.
Uncertainties in particle position cause the measurements to fall
on the vertices of a polygon that is deformed, as compared to
Fig.~1(a) without particle-position uncertainties. Here we are
interested primarily in errors in the velocity ${\bf V}$. In
Fig.~1(b), the length of the edges of the distorted polygon is an
indication of velocity. Comparing this thick irregular polygon to
the original ideal circle, it is obvious that the velocity vector
will have errors in its direction. The magnitude of the velocity
will also have errors, as we discuss next.

\subsection{C.~Illustration of PTV velocity errors for a particle in uniform circular motion}

To demonstrate the combined effect of both sources of error in
velocity, we present a simple simulation. A single particle is
assumed to perform uniform circular motion with a radius $R$ and
period $T$. A time series of a particle's $x$ and $y$ coordinates
is recorded at intervals $\Delta t$. The simulation duration is
1000 circular periods. To simulate a measurement error, we add a
random error chosen from a Gaussian distribution to the true
position. The probability of an error $x_{\rm err}$ in this
Gaussian distribution is $\propto {\rm exp}[-x_{\rm err}^2/(2
\delta_x^2)]$, and the same for the $y$ direction. We then
calculate a time series ${\bf V}(t)$, for the particle's velocity
using the simple two-frame tracking method,
Eq.~(\ref{2-frame-velocity}). We use ${\bf V}(t)$ to calculate a
time series for the kinetic energy, ${\rm KE}(t) = m |{\bf
V}(t)|^2/2$, where the mass $m$ will cancel in our final results.
We average over the entire time series yielding $\langle {\rm KE}
\rangle$, which we compare to the true kinetic energy
${\rm{KE}}_{\rm true}$ in Fig.~2. We indicate the discrepancy
between $\langle {\rm KE} \rangle$ and ${\rm{KE}}_{\rm true}$ as a
single data point in Fig.~2. We vary the size of the position
error $\delta_x$ and the sampling interval $\Delta t$, yielding
all the data points shown in Fig.~2.

Figure~2 shows the total error in $\langle {\rm KE} \rangle$ due
to both sources combined. For $\delta_x = 0$, Fig.~2 also shows
the error due to acceleration only, as indicated by circles. This
error due to acceleration is always negative, and it is most
severe for large $\Delta t$, i.e., for a slow frame rate. The
total error in $\langle {\rm KE} \rangle$, however, can be either
positive or negative, as shown by the other data points in Fig.~2.
The total error is positive at small $\Delta t$ due mainly to the
contribution of particle-position uncertainty, and it becomes
negative at large $\Delta t$ due mainly to acceleration. In
between, the total error in $\langle {\rm KE} \rangle$ has its
smallest magnitude, which for this simulation occurs at at a
sampling interval of about 3\% of the circular period, i.e.,
$\Delta t / T \approx 0.03$. This observation suggests that it may
be possible to choose a best frame rate to minimize errors, as we
discuss next.

In an actual experiment, one cannot independently measure the two
sources of error in velocity, or in quantities such as $\langle
{\rm KE} \rangle$ that are calculated from the velocity. In most
physical systems, the true motion of one particle will differ from
that of another particle, and their accelerations will vary with
time, unlike the idealized circular motion simulation in Fig.~2.
Nevertheless, examining Fig.~2 suggests a possible scheme for
choosing a $\Delta t$ that has some promise to reduce the total
error.

\subsection{D.~Scheme for reducing total error}

The scheme we suggest for reducing the total error for a quantity
(such as $\langle {\rm KE} \rangle$) computed from velocities is
as follows. The experimenter can record motion at a high frame
rate, and then analyze the data not only at the frame rate, but
also at larger $\Delta t$ by skipping one frame to double $\Delta
t$, two frames to triple $\Delta t$, and so on. Plotting the
average of KE vs. sampling interval $\Delta t$ will yield a graph
similar to Fig.~2. If the graph exhibits a nearly {\it flat spot}
at a particular value of $\Delta t$, that value is a candidate to
consider as the best value for computing KE. In Sec.~III~B, we
will examine actual experimental data as a demonstration.

\section{III.~PTV velocity errors in experiments}

\subsection{A.~Experiment}

Various experiments using PTV can have much in common, even when
the physical systems being studied are completely different.
Particle positions are measured in images that correspond to
individual video frames, and to do this they must be spaced
sufficiently that they can be distinguished. In a granular media
experiment, the particles might fill about half the
volume~\cite{Jain:02}, while in a dusty plasma experiment they are
more widely spaced, filling typically one millionth of the total
volume. There are several choices for image analysis methods to
measure the particle position, which have various advantages for
different physical systems; for our dusty plasma we use the moment
method of~\cite{Feng:07}. What might differ the most, from one
physical system to another, is the nature of the true particle
motion. In granular media, particles interact as hard spheres,
with large accelerations during the brief collision events. Dust
tracers in fluid mechanics experiments represent the opposite
extreme: they do not collide at all, but are merely swept along
with the flow of the fluid. Between these two extremes is dusty
plasma, with particles that collide softly with a screened Coulomb
repulsion, so that they undergo a gradually changing acceleration.
For recording videos of physical systems where the particles do
collide, it is useful to characterize a typical time scale. A
typical time scale is $\approx 10^{-2}~{\rm s}$ between
hard-sphere collisions in granular flow~\cite{Jain:02}. The
duration of a Coulomb collision in dusty plasma is $\approx
10^{-1}~{\rm s}$~\cite{Konopka:00}.

We carry out a demonstration experiment~\cite{Feng:10} using PTV.
Polymer microspheres (which we will refer to as dust particles) of
8.1-${\rm \mu}$m diameter are immersed in a partially-ionized
argon gas under vacuum conditions. This four-component mixture
(micron-size dust particles, rarefied gas atoms, electrons and
positive ions) is called a dusty plasma. The dust particles are
electrically charged, and they are levitated against the downward
force of gravity by a vertical electric field in the plasma,
Fig.~3(a,b). Due to their mutual Coulomb repulsion, the dust
particles tend to arrange themselves so that they are separated by
a distance $b \approx 0.7~{\rm mm}$, which is much larger than
their diameter. Using a laser heating method~\cite{Nosenko:06,
Feng:10}, the kinetic energy of dust particles is augmented, so
that their velocities fluctuate with a typical magnitude of
$\approx 1~{\rm mm/s}$, as we will find in Sec.~III~B. The dust
particles always remain within a single horizontal layer, which is
well suited for imaging.

This horizontal layer of dust particles is imaged by a high-speed
camera viewing from above. We use a 12-bit Phantom v5.2 camera,
with $1152 \times 720$ pixel resolution, and a Nikon 105 mm
focal-length macro lens. The dust particles are illuminated by a
horizontal sheet of $576~{\rm nm}$ laser light, and the camera
lens is fitted with a filter to block unwanted light at other
wavelengths. The camera's $36.2 \times 22.6~{\rm mm}^2$
field-of-view (FOV) includes $N \approx~2100$ dust particles. A
video is recorded as a series of bit-map images. The image of one
dust particle fills about ten pixels, Fig.~3(c). Due to the large
interparticle spacing, only about 2\% of the pixels in an image
are brighter than the background. The camera recorded a $20~{\rm
s}$ movie at 250 frames/s, i.e., a time between frames of $4~{\rm
ms}$.

We can compare our $4~{\rm ms}$ time interval between frames to
{\it two physical time scales} in the experiment, which are both
of interest in determining the velocity error. One physical time
scale is for acceleration for a Coulomb collision amongst dust
particles. This is typically $\approx 10^{-1}~{\rm s}$, based on
time required for the dust-particle velocity to change
significantly in the binary-collision experiment of Konopka {\it
et al.}~\cite{Konopka:00}. The other physical time scale (for
errors arising from particle-position uncertainty) is the time
required for a dust particle to move one pixel. A thermal velocity
of $\approx 1~{\rm mm/s}$ is a typical velocity in our experiment,
and at this speed a dust particle would require $\approx 30~{\rm
ms}$ to cross one pixel.

Images for each frame in the movie are analyzed using the method
of~\cite{Feng:07} to measure the dust particle positions. We
compute velocities using the simple two-frame tracking method,
Eq.~(\ref{2-frame-velocity}). We then calculate the kinetic
temperature and a correlation function, as we describe next.

\subsection{B.~Kinetic temperature}
We will demonstrate here, using experimental data, that PTV does
not yield a unique value for the kinetic temperature. Instead, it
yields a value that depends on the sampling interval $\Delta t$
between images that are analyzed. All the data presented below
come from the same experiment, with the camera always operated at
the same frame rate. To demonstrate the dependence on the sampling
interval, we will repeat our analysis by skipping frames.

The quantity that we test here is the kinetic temperature,
averaged over a time series. The time series for kinetic
temperature is calculated from mean-square velocity fluctuations
\begin{equation}\label{KT}
{T(t) = \frac{1}{Nk_B}\left[\sum\nolimits_{i=1}^N
\frac{m}{2}\left|{\bf V}_i(t)-\overline{{\bf
V}}(t)\right|^2\right],}
\end{equation}
where $N$ is the number of dust particles analyzed.
(Alternatively, one could calculate the temperature as a fit
parameter assuming a Maxwellian velocity
distribution~\cite{Knapek_thesis}.) We then average over the
entire time series, yielding the calculated result for the kinetic
temperature $\langle T \rangle$. In Eq.~(\ref{KT}),
$\overline{{\bf V}}(t)$ indicates a velocity averaged over all
dust particles in one frame, i.e., the center-of-mass velocity.
(The bar indicates an average over all dust particles in the FOV
in one image, while the brackets $\langle \cdot\cdot\cdot \rangle$
indicate averages over time.)

The results for the calculated kinetic temperature, as the
sampling time interval $\Delta t$ is varied, are shown in Fig.~4.
We see that $\langle T \rangle$ decreases gradually as the
sampling time interval increases. The error in kinetic temperature
is the difference of $\langle T \rangle$, which varies with
$\Delta t$, and an unknown true value, which does not vary. Thus,
it is clear that PTV does not yield a unique value for the kinetic
temperature, but instead a value with an error that depends on the
experimenter's choice of $\Delta t$. Moreover, the error is not
necessarily reduced by choosing a faster frame rate and using
every frame without skipping frames; if one did this, the result
for $\langle T \rangle$ would become steadily larger as the frame
rate is increased.

Having observed experimentally that there must be an error in
$\langle T \rangle$ that depends on the experimenter's choice of
$\Delta t$, we ask how the experimenter should choose $\Delta t$.
Motivated by Sec.~II~D, we will use our flat-spot scheme with our
graph of $\langle T \rangle$ vs. $\Delta t$, Fig.~4.

Examining Fig.~4, we see a general trend of $\langle T \rangle$
diminishing with the sampling time interval $\Delta t$. For the
smallest $\Delta t$ shown, the curve has a steep slope, while for
larger $\Delta t$ there is what appears to be a nearly flat spot.
Comparing to Fig.~2 for ideal circular motion, we interpret the
steep slope at small $\Delta t$ as an indication that the error is
dominated by particle-position uncertainties, which is the source
of error that becomes worse with higher frame rates. In the nearly
flat spot of Fig.~4 for our experiment, we would choose $\Delta t
\approx 0.03~{\rm s}$, corresponding to $\langle T \rangle \approx
2.5 \times 10^4~{\rm K}$ and thermal velocity is $1.3~{\rm mm/s}$.

A limitation of this scheme for choosing $\Delta t$ is that the
dust particles in the experiment do not undergo the same
acceleration all the time, unlike the simple circular-motion that
is simulated in Sec.~II~C. Therefore, there is no strong reason to
expect that the nearly flat spot in the graph of $\langle T
\rangle$  vs. $\Delta t$ will be the same as in Fig.~2 for the
circular-motion simulation. Nevertheless, this seems to be the
best guidance available to us, given what we know about the errors
in ${\bf V}$ and therefore $\langle T \rangle$ that arise from
particle-position uncertainty and acceleration.

In addition to the flat-spot scheme described above, we can
suggest an alternate scheme which leads to the same choice of
$\Delta t$ in the case of our experiment. As described in
Sec.~III~A, there are two physical time scales that can be
compared to $\Delta t$. As {\it an upper limit} for $\Delta t$, to
avoid excessive velocity errors arising from acceleration, it is
desirable to choose $\Delta t$ significantly smaller than the time
scale for velocity to change significantly in the physical system.
As {\it a lower limit} for $\Delta t$, to avoid large errors
arising from particle-position uncertainty, $\Delta t$ should be
comparable to, or larger than, the time required for a particle at
a typical velocity (in our experiment the thermal velocity) to
move one pixel. Our reasoning behind the lower limit is that the
particle-position uncertainty is generally a certain fraction of
one pixel~\cite{Feng:07, Knapek_thesis}, so that if the
displacement of a particle in $\Delta t$ is much smaller than one
pixel a large velocity error arising from particle-position
uncertainty should be expected. For this experiment, the upper
limit due to acceleration is about $100~{\rm ms}$, while the lower
limit due to particle-position uncertainty is about $30~{\rm ms}$.
Choosing $\Delta t$ significantly smaller than the upper limit,
and comparable to the lower limit, leads us to choose the same
value as the flat-spot scheme, $\Delta t \approx 0.03~{\rm s}$ in
our experiment.

We note that another approach to reduce the effect of
particle-position uncertainty was reported by Knapek {\it et
al.}~\cite{Knapek:07}. In a dusty plasma experiment similar to
ours, they operated a camera at 500 frames/s. Before carrying out
computations to track the particles and compute the kinetic
temperature, they superimposed three consecutive bit-map images,
averaging them pixel by pixel. This averaging of images has the
effect of reducing particle-position uncertainty, with the
tradeoff of a reduction of time resolution. In our scheme, instead
of averaging information while it is still in the form of bit-map
images, we focus our efforts on the particle-tracking algorithm
and choosing the time interval.

\subsection{C.~Transverse current autocorrelation function}
Aside from the kinetic energy and related quantities like the
kinetic temperature, there are other quantities one might wish to
compute from velocity data produced by PTV. For a correlation
function computed from experimental velocity and position data, we
will demonstrate here that artifacts can arise at high frame rates
that are different from the errors that appear in calculations of
the kinetic temperature. We will also demonstrate that reducing
these artifacts can require a scheme different from the one
described in Sec.~II~D for kinetic temperature.

The correlation function we will choose is the transverse current
autocorrelation function $C_T(t)$, which is used in the study of
shear motion and viscoelasticity~\cite{Balucani:00, Hu:07,
Feng:10}. This correlation function is computed from records of
particle positions $x_i(t)$ and velocities $v_{y,i}(t)$ in
orthogonal directions,
\begin{equation}\label{TCAF}
{C_T(t) = \langle j^*_y(k,0)\,j_y(k,t)\rangle / \langle
j^*_y(k,0)\,j_y(k,0)\rangle,}
\end{equation}
where
\begin{equation}\label{TC}
{j_y(k,t)=\sum\nolimits^N_{i=1} v_{y,i}(t)\,{\rm exp}[ikx_i(t)].}
\end{equation}
The quantity $j_y(k,t)$ is called a ``transverse current,'' even
though it has nothing to do with electric current. Both the
current and the correlation function $C_T(t)$ depend on a
parameter $k$ that is adjustable. In the study of
viscoelasticity~\cite{Feng:10}, values of $k$ comparable to $3/b$
are typical, where $b$ is a mean interparticle distance. (In our
experiment, $b \approx 0.7~{\rm mm}$.) Like many autocorrelation
functions, $C_T(t)$ has an initial decay at small time $t$; this
initial decay is of great interest in the study of
viscoelasticity~\cite{Feng:10}.

Using particle velocity and position data from our experiment, we
find an undesired artifact in the initial decay of the calculated
$C_T(t)$, which can be seen in Fig.~5(a). This artifact is most
prominent when the camera frame rate is high and when we use the
simple two-frame tracking method to prepare the velocity data used
in calculating $C_T(t)$. We attribute this artifact to
particle-position uncertainty.

Although this artifact arises due to a high camera frame rate, it
is unattractive to attempt to eliminate it by reducing the frame
rate. Many data points during the initial decay are required for
the study of viscoelasticity, and this requires a high frame rate.
Thus, we need a different approach to reduce the artifact, like
the two-step approach we describe next, for reducing the effects
of particle-position uncertainty.

In the first step, we use a three-frame tracking method, which
reduces the artifact as seen in Fig.~5(b). In this three-frame
tracking, we do not entirely skip any frames, because that would
reduce the temporal resolution of $C_T(t)$. Instead, we interlace
pairs of frames, and calculate the positions and velocities as:
\begin{equation}\label{3-data-position}
{{\bf X}_{j} = ({\bf x}_{j-1}+{\bf x}_{j}+{\bf x}_{j+1})/3,}
\end{equation}
\begin{equation}\label{3-data-velocity}
{{\bf V}_{j} = ({\bf x}_{j+1}-{\bf x}_{j-1})/2\Delta t.}
\end{equation}
Here, $x_{j-1}$, $x_{j}$, $x_{j+1}$ are the positions of a
particle in any three consecutive frames. Equations
(\ref{3-data-position}) and (\ref{3-data-velocity}) can be derived
from the linear regression for three data points. Using three
frames rather than two requires a sufficiently fast frame rate,
which is satisfied in our experiment with a time interval between
frames of $4~{\rm ms}$, which is much shorter than the time scale
for changes in the dust particle's velocity $\approx 10^{-1}~{\rm
s}$. (While this three-frame tracking method has the advantage of
reducing the unwanted artifact in the initial decay of $C_T(t)$,
it may also have an effect on the long-time oscillations of
$C_T(t)$, which we have not studied.)

To further reduce the artifact, in the second step, we smooth
$j_y(k,t)$ before calculating $C_T(t)$. This averaging requires a
sufficiently high frame rate. We smooth the transverse current
$j_y(k,t)$ with a moving average over five consecutive frames
before calculating $C_T(t)$. The result for $C_T(t)$, Fig.~5(c),
is a great reduction of the artifact without an unnecessary loss
of time resolution in the initial decay.

To verify that these two steps reliably remove artifacts without
changing the true shape of the initial decay, we carried out
numerical simulations, where the true shape is known. These
simulations used a Langevin molecular dynamic
method~\cite{Feng:10} to track particles by integrating their
equations of motion. To mimic the experiment's particle-position
uncertainty, we added random errors to the positions in the
simulation as in Sec.~II~C. The simulation results, not shown
here, confirm that the artifact is diminished by the two steps
described above, and they also confirm that this improvement
results in only a negligible change in the initial decay of the
calculated $C_T(t)$, as compared to the true $C_T(t)$ without the
random errors.

\section{IV.~Summary}
In summary, we have studied two sources of errors in PTV using
high-speed cameras. What may surprise some users is that a higher
frame rate is not necessarily better, and in fact it can greatly
worsen velocity errors arising from particle-position uncertainty.
We described our solutions to reduce these errors: two schemes
that lead to the same choice for $\Delta t$ when calculating
kinetic temperature, and a two-step approach for computing an
autocorrelation function. We demonstrated these solutions using
the data from a dusty plasma experiment. This work was supported
by NSF and NASA.

\begin{figure}[p]
\caption{\label{sketch} (Color online) Simulations of particle
trajectories for circular motion (a) with no particle-position
uncertainty, and (b) with particle-position uncertainty. The true
particle positions at time intervals $\Delta t$ are marked by
circular dots. The triangles in (b) indicates particle positions
including an error from a Gaussian distribution with an
uncertainty $\delta_x$.}
\end{figure}

\begin{figure}[p]
\caption{\label{KE} (Color online) Velocity errors for a simulated
particle executing uniform circular motion with period $T$, as in
Fig.~1(b). The velocity is calculated using
Eq.~(\ref{2-frame-velocity}), and then the kinetic energy is
calculated and averaged over 1000 cycles of the circular motion.
The true kinetic energy is $m (2 \pi R / T)^2/2$. The severity of
the two sources of error in the kinetic energy depends on the
sampling interval $\Delta t$. The dominant error arises at small
$\Delta t$ from particle-position uncertainty, and at large
$\Delta t$ from acceleration. There is a nearly flat spot at
$\Delta t/T \approx 0.3$, where the total error from two sources
is smallest.}
\end{figure}

\begin{figure}[p]
\caption{\label{Image} (Color online) Configuration for the
experiment. (a) Dust particles were levitated in a horizontal
layer above a lower electrode. A vacuum chamber, not shown here,
was filled by a weakly-ionized plasma, consisting of electrons,
positive argon ions, and a neutral argon gas. A pair of 532-nm
laser beams accelerated the dust particles to raise the kinetic
temperature. A top-view camera imaged the dust particles, which
were illuminated by a horizontal sheet of 576-nm laser light. (b)
Dust particles are electrically levitated in a horizontal layer.
(c) Experimental bitmap image from one frame of a video. Only
$\approx 1/60$ of the experimental image is shown. The inset is a
magnified image of one dust particle.}
\end{figure}

\begin{figure}[p]
\caption{\label{KTfig} Experimental data for the kinetic
temperature, calculated using Eq.~(\ref{KT}). To vary the sampling
time interval $\Delta t$, we skipped frames in the data analysis.
We note the appearance of a nearly flat spot, similar to the one
seen in Fig.~2 for the circular-motion simulation. In Sec.~III~B
we suggest that for calculating kinetic temperature, it is
desirable either to choose $\Delta t$ in the flat spot of this
graph, or to choose $\Delta t$ comparable to the time required for
a particle at the thermal velocity to move a distance of one
pixel. Both of these schemes lead to the same choice in this
experiment, $\Delta t \approx 0.03~{\rm s}$.}
\end{figure}

\begin{figure}[p]
\caption{\label{TCAF} Experimental data for $C_T(t)$, an
autocorrelation function of the time series $j_y(k,t)$, as defined
in Eq.~(\ref{TCAF}). An unwanted artifact appears in the initial
decay in (a) with two-frame tracking, i.e., using
Eq.~(\ref{2-frame-velocity})) to calculate velocities. This
artifact is reduced in (b) by using three-frame tracking, i.e.,
using Eqs.~(\ref{3-data-position}) and (\ref{3-data-velocity}). It
is further reduced in (c) by smoothing the the time series
$j_y(k,t)$ before calculating its autocorrelation. Data shown here
were computed for $kb=3.5$, where $b$ is the mean interparticle
distance.}
\end{figure}

\end{document}